\documentclass[12pt,twoside]{article}
\usepackage{fleqn,espcrc1}

\usepackage{epsfig}

\usepackage[figuresright]{rotating}

\newcommand{\AmS}{{\protect\the\textfont2
  A\kern-.1667em\lower.5ex\hbox{M}\kern-.125emS}}

\title{Neutral Pion Distributions in PHENIX at RHIC}

\author{G. David\address[BNL]{Brookhaven National Laboratory, 
        Physics Department, \\ 
        Upton, NY 11973, USA} \\
 	for the PHENIX Collaboration\footnote{For the full 
	PHENIX Collaboration authors list and acknowledgments, 
	see ``PHENIX Overview'' presented by W. A. Zajc in this volume.}
	}
       
\begin{document}

\maketitle

\begin{abstract}
Transverse momentum spectra for identified $\pi^0$'s 
in the range 1~GeV/c~$< p_T <$~4~GeV/c have
been measured by the PHENIX experiment in Au-Au collisions 
at $\sqrt{s}=130$ GeV.   
The spectra from peripheral nuclear collisions 
are consistent with the simple expectation of scaling the spectra 
from p+p collisions by the average number of
nucleon-nucleon binary collisions.  The spectra from central  
collisions and the ratio of central/peripheral spectra
are significantly suppressed when compared to
point-like scaling.

\end{abstract}

\section{The PHENIX Experiment}


\subsection{Physics goals}

The Relativistic Heavy Ion Collider (RHIC) at Brookhaven National
Laboratory started regular operations in June 2000, opening new
frontiers in the study of hadronic matter under unprecedented
conditions of temperature and energy density.  The research is
focused on the phase transition associated with quark deconfinement
and chiral symmetry restoration expected to take place under these
conditions.

PHENIX is one of the four experiments at RHIC.  It is designed
to cover the entire time-scale of the interaction, from initial hard
scattering to final state interactions by 
simultaneous measurement of a wide range of probes in the same detector.

Here, the results on neutral pion production in the
$1.0~<~p_T~<~4.0$ ~GeV/c transverse momentum range are presented.  
High $p_T$
hadrons are produced primarily by initial hard scattering of partons,
and their rate of production in $pp$ collisions can be calculated in 
perturbative QCD, which simply scales to nuclear collisions
by the relative number of binary nucleon-nucleon collisions.
However, in the case of colliding nuclei, there are 
several nuclear effects that can modify the spectra of hadrons
even in the ordinary nuclear medium, such as shadowing and $p_T$ broadening.  
In addition, there are predictions~\cite{Wang00,Levai00}
that if the medium becomes very
dense, the scattered partons may lose considerable energy via
gluon bremsstrahlung or ``jet quenching''.  According to 
those predictions,
this effect would cause a significant depletion of the
spectra at high $p_T$.

\subsection{Experimental setup}

The PHENIX detector consists of an axial-field magnet surrounded
by two central arms (called East and West), each one subtending
$90^0$ in azimuth and $\pm0.38$ units of pseudorapidity.
Two spectrometers at forward angles ($10^o<\theta<35^o$)
around the beam axis serve to identify and track muons.
During the Year 2000 data taking run, only a subset of the 
detectors of the central arm were read out and analyzed.

Trigger and basic event characterization was provided by
two sets of beam-beam counters (BBC) covering $2\pi$ in azimuth
and $\eta=\pm(3.0-3.9)$, as well as by two zero-degree
calorimeters located $\pm18.25$ m from the collision point
and covering $|\eta|>6$~\cite{Milov01}.  The primary interaction trigger
is generated by the coincidence between the 
beam-beam counters which detect $92\pm2$\% of the nuclear
interaction cross section of $\sigma_{int} = 7.2$ barns.  
Another trigger is generated by a coincidence between the two
zero-degree calorimeters which are sensitive to unbound spectator
neutrons from the nuclear interactions or from Coulomb dissociation.
The correlation between the BBC and ZDC signals are
used to determine the centrality of the collision.  
They are also used to establish the number of
participant nucleons ($N_p$) and binary collisions ($N_{coll}$)
using a Glauber-model calculation~\cite{Milov01}.

The central arm spectrometers consist of a multiplicity
vertex detector (MVD), drift chambers (DC), two layers of
pad chambers (PC), a gas-filled ring imaging Cherenkov detector
(RICH), a time expansion chamber (TEC), and a high granularity 
electromagnetic calorimeter (EMC) consisting of eight sectors,
each covering $|\eta|~<~0.38$ and $\Delta\phi~=~22^{\circ}$.  Six
of the calorimeter sectors are lead-scintillator sampling calorimeters
(PbSc), while two sectors consist of lead glass calorimeters (PbGl).
The data presented here are obtained from two PbSc sectors in
the West arm spectrometer, which have 
$8.2\%/\sqrt{E({\rm{GeV})}}~\oplus~1.9\%$ energy resolution
and $5.7/\sqrt{E({\rm{GeV}})}~\oplus~1.6$~mm position resolution for
electromagnetic showers in a low multiplicity environment.  
The energy calibration was established using minimum
ionizing particles, maintained with a laser monitoring system,
and verified from the data using $E/p$ matching for identified
electrons in low multiplicity events, as well as with the mass of the
$\pi^0$.  The systematic error on the overall energy scale is
less than 1.5\%~\cite{Et01}.

\subsection{Event selection, data set}

Only events taken at full magnetic field and satisfying
the primary interaction trigger, as explained above, 
are analyzed.  Additional cuts include the measured
event vertex position ($|z|<30$~cm), consistency between the
ZDC and BBC interaction time measurement.
Altogether, 1.17 million events passed these
cuts and form the sample referred to as ``minimum bias''.
Based upon the correlation of the measured BBC charge and 
ZDC energy, centrality classes were established as fractions
of the total nuclear cross section.  In this analysis
``central'' refers to the 0-10\% most central collisions,
while ``peripheral'' means the upper 60-80\% range of $\sigma_{int}$.
\footnote{At the Quark Matter conference we presented the
upper 75-92\% range of $\sigma_{int}$ as ``peripheral''.  This is
now replaced by the upper 60-80\% of $\sigma_{int}$ where the
sample is much cleaner and the results are less biased
by the inefficiencies of the trigger in the most peripheral
collisions.  The centrality selection for the ``central'' spectra
is the same (0-10\% most central), but the points and the
systematic errors have been revised.}

\section{Analysis}

Neutral pions have been measured using their 
$\pi^0~\rightarrow~\gamma\gamma$ decay mode.
In this measurement, the invariant mass of the photon pairs has
to be within a narrow ($2\sigma$) window around the observed 
$\pi^0$ mass, thus allowing for less stringent photon identification 
cuts, which reduces the systematic errors on efficiency losses 
due to these cuts.  

Each cluster found in the calorimeter is subject to a
timing cut: it has to arrive within 2.5 ns with respect to
the expected time-of-flight of a photon coming from the event vertex
(TOF cut).  This cut eliminates slow hadrons, in particular
anti-neutrons which are a major source of neutral clusters
in the 1-2 GeV energy range.  In addition,
the shape of each cluster is compared to the known and parametrized
shape of electromagnetic showers, and the $\chi^2$ 
of the difference between the observed and predicted shower shape
is calculated \cite{Et01}.  A  $\chi^2<3$ cut is then applied to the showers.
Both cuts are designed to keep the photon efficiency as high as
possible.
Therefore the accepted clusters have a significant contribution for other
particles which is removed by the backround subtraction method 
described below.
Furthermore, the efficiency
of both cuts depends on the event multiplicity.  This was
verified from the data by comparing $\pi^0$ peak contents
at a given $p_T$ extracted with different photon identification
cuts, as well as by studying the effect of the cuts on well
identified electrons.

The $\gamma\gamma$ invariant mass is calculated from all pairs of 
clusters in an event 
passing the photon identification cuts.  The combinatorial
background is estimated using an event mixing method, which
after proper normalization, is subtracted from the invariant
mass distribution.  

The $\pi^0$ reconstruction efficiency is calculated by the 
following procedure.
First, the effect of a second particle contributing to the
same cluster is investigated (overlaps).  Simulated single 
electromagnetic showers are merged both in real and simulated events.  
For different single shower energies and event centralities, the
distributions of the ratios of the measured and original 
energies are stored (referred to as photon energy and event 
centrality-dependent $f(E_{\gamma},cent)$ ``smearing'' functions).  These
functions are then used in a fast
Monte Carlo simulation as follows.  Neutral pions are generated with the
expected $p_T$ distribution and allowed to decay.  For those cases, when
both decay photons reach the calorimeter, their respective
energies are randomized with the appropriate $f(E_{\gamma},cent)$
smearing function, and the invariant mass is calculated using the
randomized energies.  The resulting simulated line-shapes
are compared to line-shapes obtained from the data after
mixed event subtraction. They agree very well, and the same cuts
are applied to the data and the simulations to establish the
efficiency.

Simulations are also used to determine the background from
particles striking the pole-tips and structural elements of detectors
in front of the calorimeter.  
An additional source of background arises from those $\pi^0$'s 
produced close to (but not at) the collision vertex which 
reconstruct in the calorimeter
with the proper invariant mass, increasing the 
true $\pi^0$ yield.  This background
is also estimated using 
simulations (HIJING 1.35~\cite{Gyulassy93}).  
The calculated contribution of non-vertex but
properly reconstructed $\pi^0$ is $\sim8\%$ at $p_T=1~GeV/c$
and gradually decreases to $\sim6\%$ at $p_T>2.5~GeV/c$.  This yield
has been subtracted from the measured $\pi^0$ yield.

\section{Results and discussion}

\begin{figure}[htb]
\begin{flushleft}
\begin{minipage}[t]{75mm}
\epsfig{file=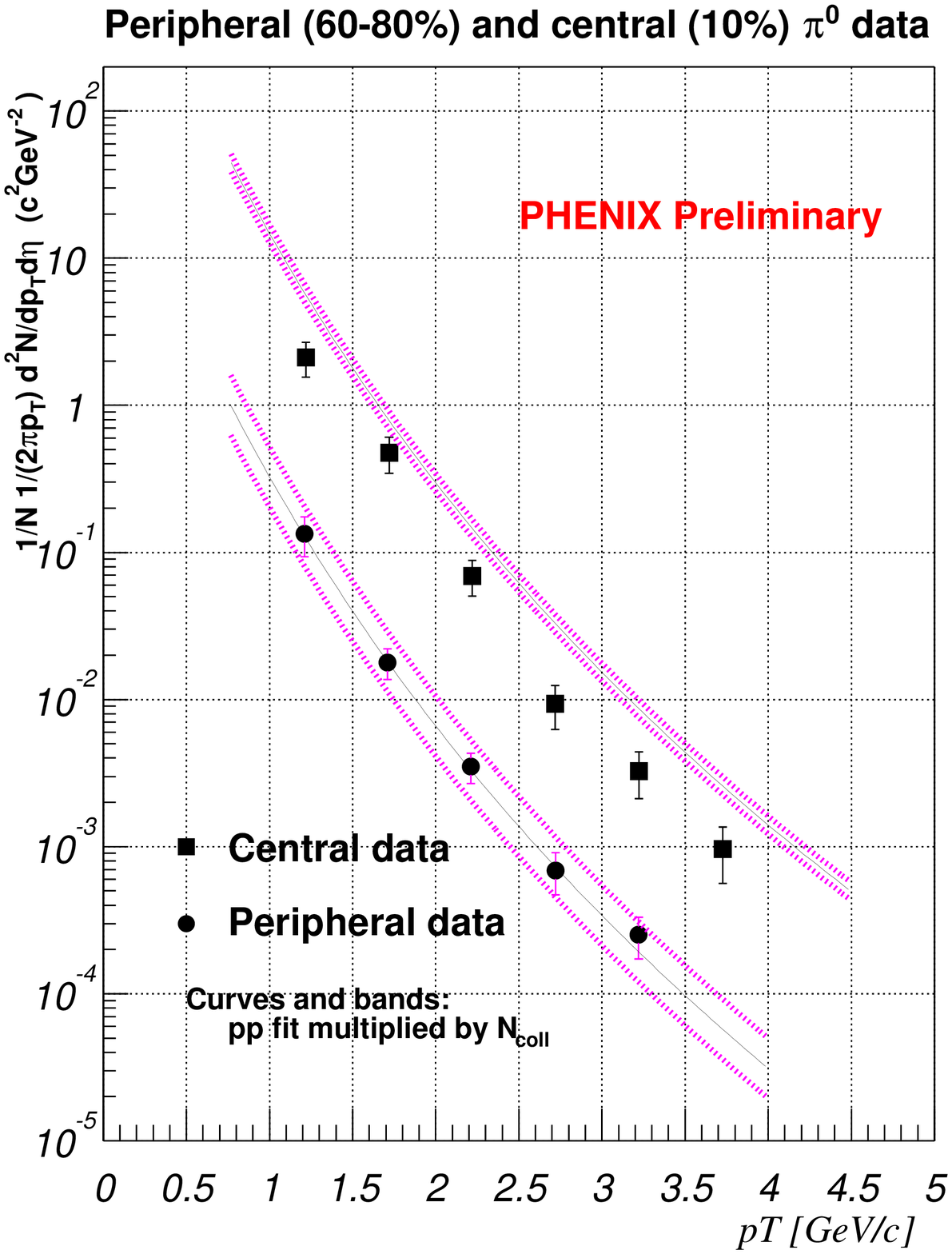,width=7.5cm,angle=0}
\caption{Semi-inclusive $\pi^0$ $p_T$ distribution
	$(1/N_{int})(dN_{\pi^0}/2\pi p_Tdp_Tdy)$)
	 in the upper 60-80\% peripheral events (solid
	circles) and the 10\% most central events
	(solid squares).  The lines are a parametrization
	of $pp$ charged hadron spectra, scaled by the
	mean number of collisions $N_{coll}$.  The bands indicate
	the possible range due to the systematic error on $N_{coll}$.
	}
\label{fig:plot_pt_ua1}
\end{minipage}
\end{flushleft}
%
\vspace{-16cm}
\begin{flushright}
\begin{minipage}[t]{75mm}
\epsfig{file=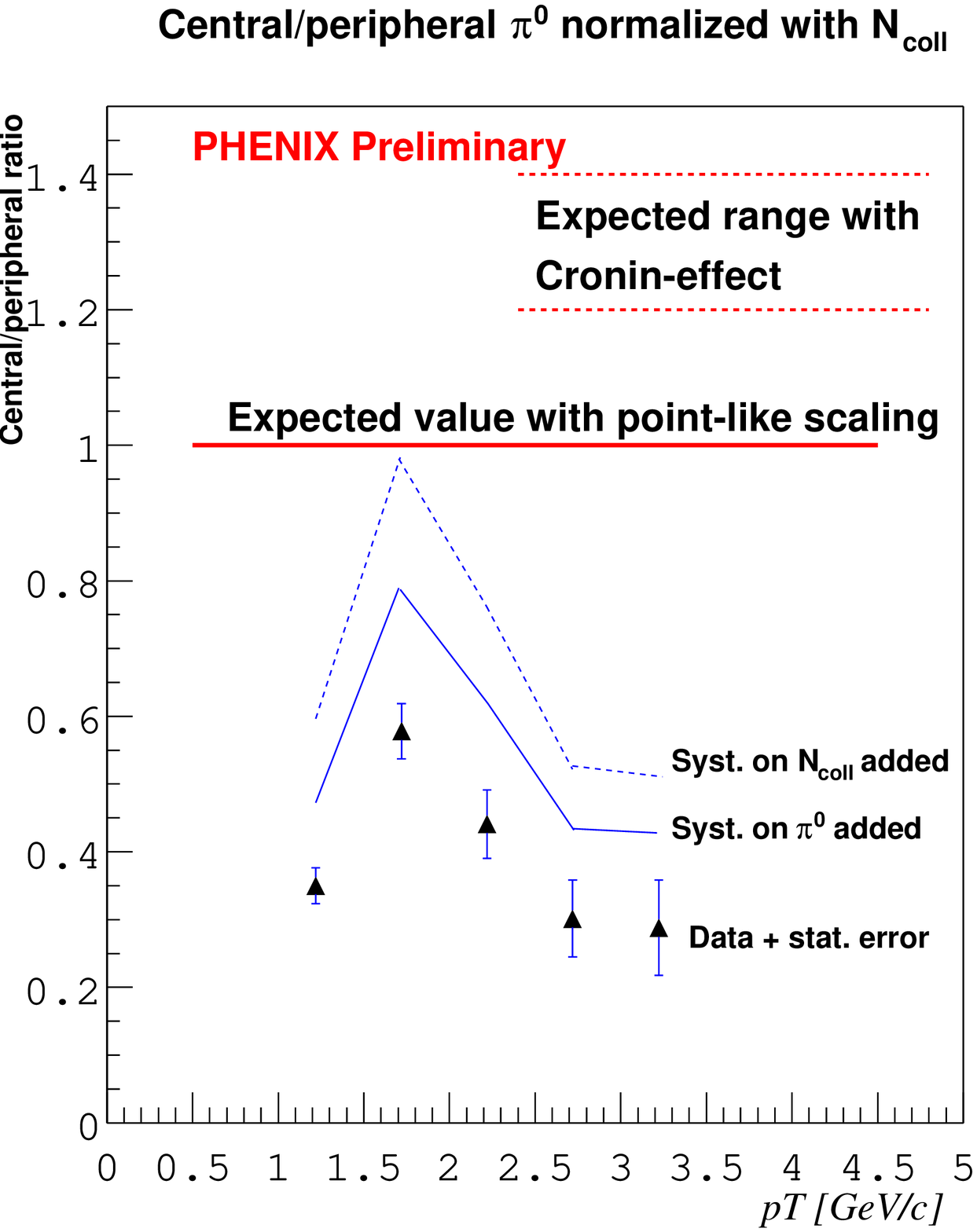,width=7.5cm,angle=0}
\caption{Ratio of the central and peripheral $\pi^0$ spectra,
	both normalized by the mean number of collisions
	(857 and 19, respectively).  Error bars are statistical
	only.  Solid line: systematic errors of the
	$\pi^0$ measurement added in quadrature.  
	Dashed line: systematic errors on 
	the number of binary collisions added in quadrature.
	Light shaded area: expected range of the ratio
	with the Cronin-effect.
	}
\label{fig:ratio_centperiph2}
\end{minipage}
\end{flushright}
\end{figure}

The semi-inclusive transverse momentum distribution of $\pi^0$ in peripheral
(upper 60-80\% of $\sigma_{int}$) and 10\% most central 
Au+Au collisions
is shown in Figure~\ref{fig:plot_pt_ua1}.  At high $p_t$ the
peripheral spectrum is limited by statistics.  Error bars include
both statistical and systematics errors. 

Both spectra are compared to $p_T$ spectra derived from nucleon-nucleon
data.  Since there is no measurement of $\pi^0$ production in $pp$ at
$\sqrt{s}=130$~GeV, this reference spectrum is derived from
UA1~\cite{UA1} and CDF~\cite{CDF} charged hadron spectra.  First, the 
available data are fitted with a function 
$d\sigma/2\pi p_T dp_T dy = A / (p_0+p_T)^n$, then
the fit parameters $A,p_0,n$ are interpolated to RHIC energy.
The result is divided by $\sigma_{pp}=42$ mb for the yield and by 
1.6 to obtain the
pion content from the unidentified charged spectra.
(The actual parameter values are $A=275000/42/1.6$, $p_0=1.71$ and
$n=12.42$.)  This parametrized curve is then multiplied by the
estimated mean number of binary nucleon-nucleon collisions 
($19\pm 11$ and $857 \pm 128$
in peripheral and central collision, respectively).  The systematic
error on the number of collisions is indicated by the two bands.

In peripheral collisions the scaled $pp$ parametrization describes the
results very well, but it significantly overpredicts the measured spectrum
in central collisions, particularly at higher $p_T$.  The observed
deficit in the $\pi^0$ yield is even more surprising if one takes
into account that at 3-4 GeV/c a Cronin-type enhancement 
due to $p_t$ broadening above the scaled $pp$ 
distribution would be expected.

The same deficit can be seen in Figure~\ref{fig:ratio_centperiph2}
without referring to parametrized $pp$ results.  Both the central
and the peripheral spectra are normalized by the respective number
of binary collisions, then divided point-by-point.
The central/peripheral ratios are shown as triangles, and the error bars
are statistical.  The solid line gives the  upper limit on the ratio 
if the systematic errors of the $\pi^0$ spectra are added in quadrature.
The dashed line adds (in quadrature) the systematic error
on the number of collisions ($N_{coll}$).  The central/peripheral
ratio, normalized by $N_{coll}$, is expected
to be one in the case of simple scaling with
$N_{coll}$.  However, the measured ratio is much smaller. 
The expected range of the central/peripheral
ratio with a Cronin-effect included is also shown.

\begin{figure}[htb]
\epsfig{file=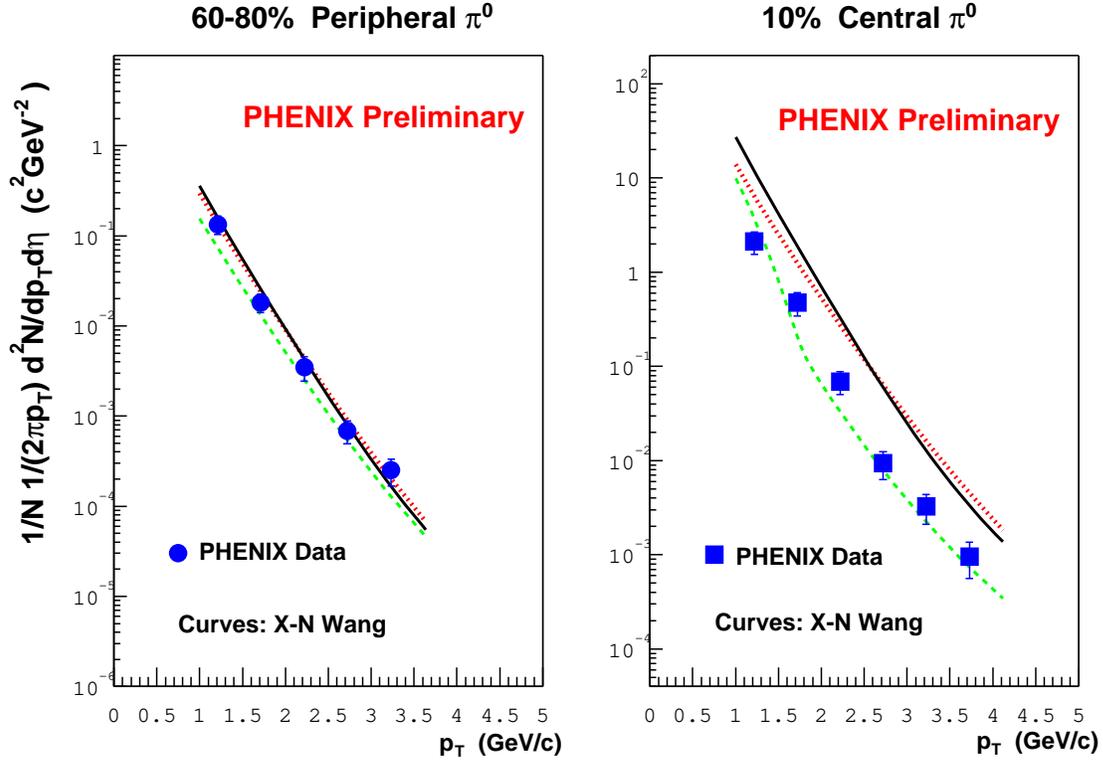,width=15.0cm,angle=0}
\caption{Comparison of PHENIX $\pi^0$ spectra to theoretical
	calculations under three scenarios and for two centralities.
	The points are the same as on Figure~\ref{fig:plot_pt_ua1}.
	The curves are calculations of X-N. Wang~\cite{Wang00}.
	Solid lines are a pQCD calculation for $pp$ scaled by the mean
	number of binary collisions.  The dotted lines add shadowing
	and $p_T$ broadening.  The dashed lines add a 
	$dE/dx=0.25$~GeV/fm parton energy loss.
	}
\label{fig:ptdist_wang_cent10}
\end{figure}

Figure~\ref{fig:ptdist_wang_cent10} shows the results for both the
peripheral and central collisions  compared to
three theoretical calculations~\cite{Wang00} (curves).  
The solid lines are
a straightforward pQCD calculation for $pp$,
with simple scaling to Au-Au collisions
by the mean number of binary collisions \cite{Wang00}.  
The dotted lines are a
calculation where effects of nuclear shadowing and $p_T$
broadening are added, and result in a change of slope, suppressing
the soft part of the spectrum and enhancing the hard scattering
part (Cronin effect).  
The calculation plotted with dashed lines
adds a constant  $dE/dx=0.25~GeV/fm$ parton energy loss to the shadowing and 
$p_T$ broadening.  
The peripheral data are consistent with all three scenarios.
However, the central data are well below the first and second
(pQCD and shadowing/Cronin) curve, but they are not inconsistent
with the third scenario that includes a parton energy loss.


\section{Conclusion}

Transverse momentum spectra for neutral pions in central and peripheral
$\sqrt{s}=130$ GeV Au+Au collisions have been presented.
The peripheral spectrum is consistent with the simple scaling
of $pp$ collisions with the mean number of binary nucleon-nucleon
collisions.  In the central spectra, a significant deficit with
respect to this point-like scaling is observed at high transverse
momenta.

\end{document}